\documentclass[%
 reprint,
 amsmath,amssymb,
 aps,
]{revtex4-2}

\usepackage{graphicx}
\usepackage{dcolumn}
\usepackage{bm}
\usepackage{subfigure}
\usepackage{booktabs}

\begin{document}

\preprint{APS/123-QED}

\title{Measurement of the $^{14}$C spectrum with Silicon Drift Detectors: \\ towards the study of forbidden $\beta$ transitions}

\author{Andrea Nava$^{1,2}$}
 \email{andrea.nava@mib.infn.it} 
\author{Leonardo Bernardini$^{3,4}$}
\author{Matteo Biassoni$^{2}$}
\author{Tommaso Bradanini$^{1,2}$}
\author{Chiara Brofferio$^{1,2}$}
\author{Marco Carminati$^{3,4}$}
\author{Giovanni De Gregorio$^{5,6}$}
\author{Carlo Fiorini$^{3,4}$}
\author{Giulio Gagliardi$^{1,2}$}
\author{Peter Lechner$^{7}$}
\author{Riccardo Mancino$^{8,9,10}$}
\affiliation{%
 {$^1$Dipartimento di Fisica G.Occhialini, Università di Milano - Bicocca, Milano, 20126, Italy} \\
 {$^2$INFN, Sezione di Milano - Bicocca, Milano, 20126, Italy} \\
 {$^3$DEIB, Politecnico di Milano, Milano, 20133, Italy}\\
{$^4$INFN, Sezione di Milano, Milano, 20133,Italy}\\
{$^5$Dipartimento di Matematica e Fisica, Università degli Studi della Campania “Luigi Vanvitelli”, Caserta, 81100, Italy}\\
{$^6$INFN, Sezione di Napoli, Napoli, 80126, Italy}\\
{$^7$Halbleiterlabor der Max-Planck-Gesellschaft, Garching, 85748, Germany}\\
{$^8$Institute of Particle and Nuclear Physics, Faculty of Mathematics and Physics, Charles University, Prague, 18000, Czech Republic}\\
{$^9$Institut f{\"u}r Kernphysik (Theoriezentrum), Fachbereich Physik, Technische Universit{\"a}t Darmstadt, Darmstadt, 64298, Germany}\\
{$^{10}$GSI Helmholtzzentrum f{\"u}r Schwerionenforschung, Darmstadt, 64291, Germany}\\
}%

\date{\today}

\begin{abstract}
The ASPECT-BET (An sdd-SPECTrometer for BETa decay studies) project aims to develop a novel technique for the precise measurement of forbidden $\beta$ spectra in the 10 keV - 1 MeV range. This technique uses a Silicon Drift Detector (SDD) as the main spectrometer, surrounded, if necessary, by a veto system to reject events with only partial energy deposition in the SDD. Accurate knowledge of the spectrometer's response to electrons is essential to reconstruct the theoretical shape of the $\beta$ spectrum. To compute this response, GEANT4 simulations optimized for low-energy electron interactions are used. In this article, we present the performance of these simulations in reconstructing the electron spectra, measured with SDDs, of a $^{109}$Cd monochromatic source, both in vacuum and in air. The allowed $\beta$ spectrum of a $^{14}$C source is also measured and analyzed, and it is shown that the experimental shape factor commonly used in the literature to reconstruct the measured spectrum is not necessary to explain the spectrum.
\end{abstract}

\maketitle


\section{Introduction}\label{sec1}

Nuclear theories are key components in the interpretation of several results in neutrino physics, as in the case of the $0\nu\beta\beta$ decay search \cite{Barea2013} or the reactor oscillation experiments \cite{Hayes2014}. However, the lack of a single model capable of accurately predicting all experimental observables results in significant nuclear-related systematics.\\
The predicted shape of the forbidden $\beta$ spectra is highly dependent on the theoretical description and can be an important tool for discriminating between different nuclear models \cite{Kostensalo2017}. In addition, measuring different isotopes with the same setup would increase the ability to rule out theories that cannot predict all spectra within the same framework.\\
To this end, unprecedentedly precise measurements have recently been made for some forbidden $\beta$ spectra using various technologies. So far, competitive experiments at room temperature have been realized only for beta-active isotopes naturally present in the detector material itself, as in scintillators (e.g. $^{176}L$u in LSO:Ce or LuAG:Pr \cite{Quarati2023}) or in semiconductors (as $^{113}$Cd in CdZnTe \cite{Cobra2020}). 
The best results in energy threshold and resolution have been achieved with scintillating low-temperature detectors, such as LiInSe$_2$ \cite{Leder2022} and LiI \cite{Pagnanini2023} for $^{115}$In, and with metallic magnetic calorimeters (MMCs) used to measure forbidden $\beta$ spectra of $^{36}$Cl \cite{Rotzinger2008}, $^{151}$Sm \cite{Kossert2022} and $^{99}$Tc \cite{Paulsen2023}. \\
MMCs are of particular interest because they can embed the beta-active isotope in the thermal absorber (a very thin gold foil), thus satisfying the requirement to measure different isotopes with the same setup in a relatively simple way \cite{Loidl2019}. However, they have all the technical drawbacks of working at very low temperatures with dilution refrigerators. Isotopes with short half-lives cannot be studied because it always takes a long time to reach the base temperature. In addition, the strict requirement for absorber dimensions also limits the maximum mass of the isotope that can be safely embedded, discouraging the use of MMCs with extremely rare isotopes and/or extremely long half-lives. \\
\\ 
ASPECT-BET (An sdd-SPECTrometer for BETa decay studies) is a project that aims to measure a set of forbidden $\beta$ spectra using Silicon Drift Detectors (SDDs) \cite{Biassoni2023}. These detectors are currently being developed for electron spectroscopy in the context of TRISTAN, the upgrade of the KATRIN detector to search for keV sterile neutrinos \cite{Mertens2019}. \\
SSDs are suitable candidates for precise $\beta$ spectroscopy due to several advantages. SDDs can operate at room temperature with good energy resolution ($\sim$200 eV at 5.9 keV) and can sustain a high interaction rate \cite{Mertens2021}. Operating at room temperature also allows the measurement of short half-life isotopes, such as those produced in reactors, making this technique complementary to cryogenic techniques. SDDs can be used in conjunction with auxiliary veto detectors, such as scintillators, to reject events where only a fraction of the total energy is deposited in the main detector. Finally, the complete decoupling of the SDD spectrometer from the source simplifies the use of the same setup to study different isotopes. \\  
\\ 
Spectroscopy of electrons with energies in the 10 keV - 1 MeV range from an external radioactive source presents several challenges, the most important are:
\begin{itemize}
    \item Energy loss in the source, especially if it is thick or encapsulated (self-absorption);
    \item Energy loss in the dead layer, usually present in all silicon devices;
    \item Incomplete energy deposition due to electron backscattering after impinging on the detector;
    \item Incomplete energy deposition due to escape of characteristic X-rays and bremsstrahlung;
    \item Incomplete charge collection if the event is too close to the detector boundary.
\end{itemize}
Our goal is to develop a complete model of the system, including both the source and the detector, based on GEANT4 simulations \cite{Agostinelli2003} and analytical descriptions of detector non-idealities, that can account for all these effects in order to accurately predict the shape of electron spectra measured with an SDD. \\
Some of these parameters, such as the energy loss in the dead layer or the electron backscattering, have already been studied in the KATRIN context \cite{GUGIATTI2020164474, Biassoni2020}, so our main focus is now on the source-related effects that affect the spectral shape. \\
We decided to take a stepwise approach, starting with a commercial encapsulated monochromatic electron source measured in vacuum and air. The aim is to test the ability of GEANT4 to simulate low-energy electron scattering on light materials, such as the source plastic capsule and air, and then to extend the study to a measurement with an allowed $\beta$-decay source. 

\section{Experimental setup}\label{sec2}
\begin{figure}
\centering
\begin{subfigure}{}
\includegraphics[width=0.33\textwidth]{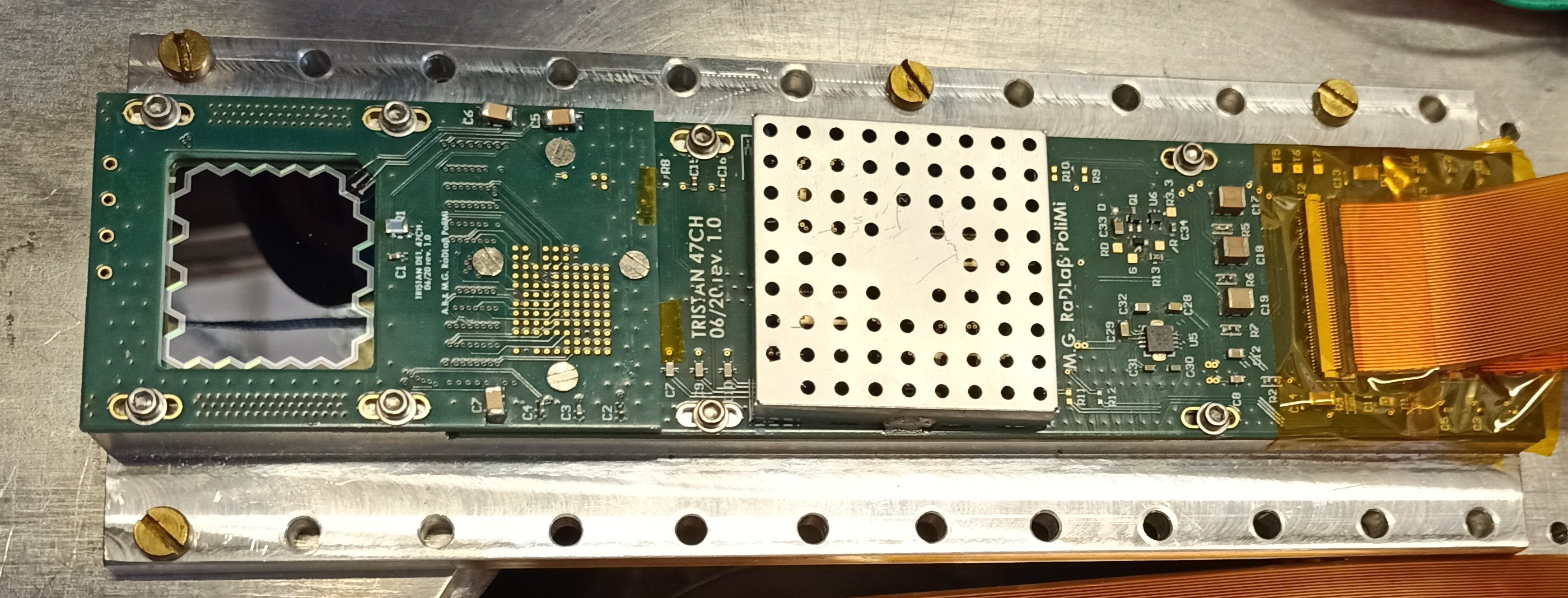}
\end{subfigure}%
\begin{subfigure}{}
\includegraphics[width=0.13\textwidth]{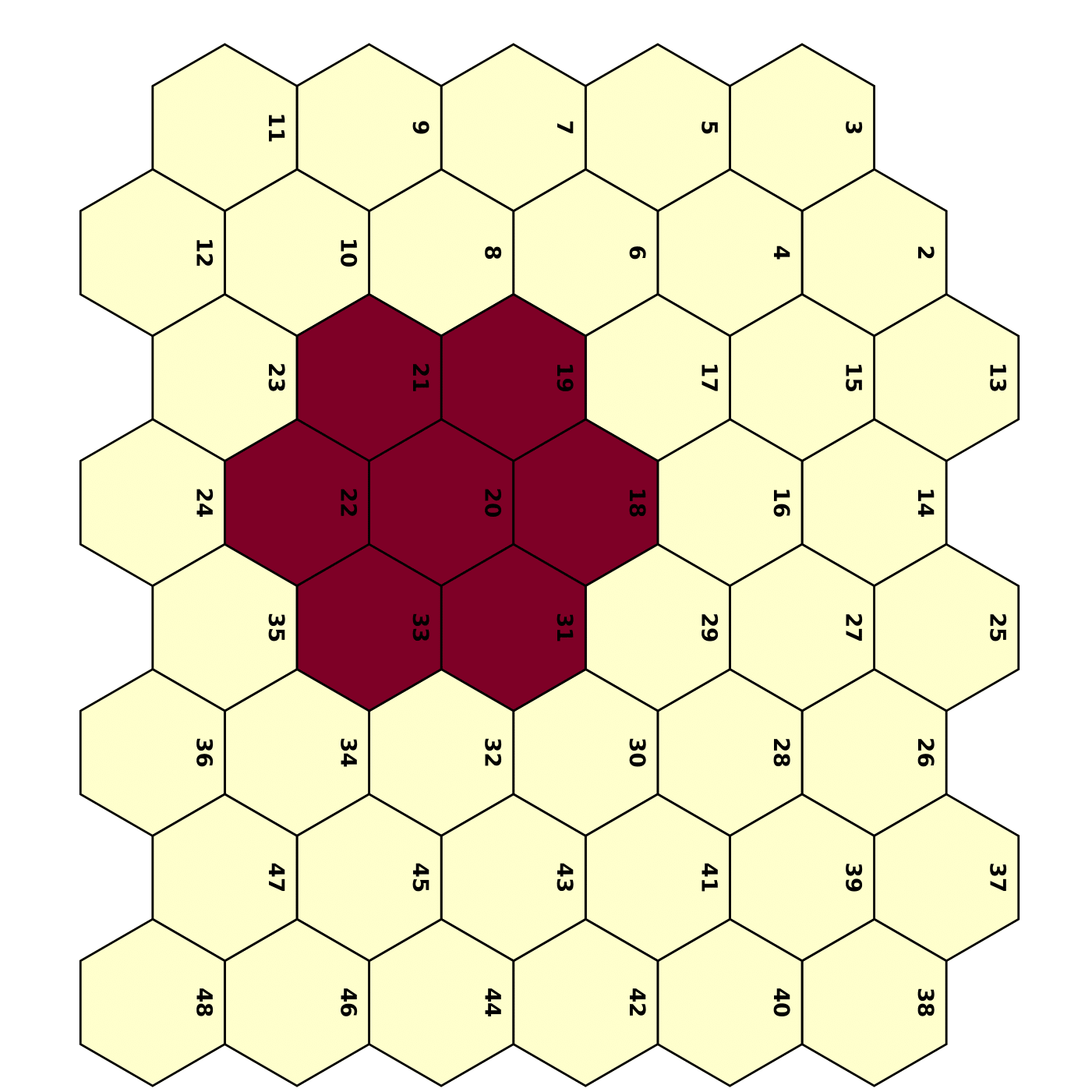}
\end{subfigure}
\caption{47-pixel SDD matrix used for all the measurements (left). Scheme of the pixels, only the ones in red were acquired (right).}\label{SDD_geom}
\end{figure} 
The silicon drift detectors we used are the same devices built at MPP-HLL (Munich, Germany) for the TRISTAN detector, which are 450 $\mu$m thick and have a hexagonal shape with a side length of 1.5 mm. Since these detectors are intended for spectroscopy of electrons with energies down to 10 keV or less, they have a thin $\sim10$ nm SiO$_2$ entrance window. \\
SDDs have been operated in a vacuum chamber where a vacuum level down to $10^{-5}$ mbar can be achieved. The amplification is done in vacuum using 12-channel Ettore ASICs \cite{IEEE}. 
In this work, we acquired a group of 7 adjacent SDDs from a 47-pixel matrix, as shown in Figure \ref{SDD_geom}, using a CAEN DT5743 digitizer. These 7 channels were chosen because they have similar energy resolutions and thresholds. The central SDD of these 7 was considered the main one, while the surrounding 6 detectors were used to reject those events with energy deposition in more than one pixel (M2, or "multiplicity-2 events"). \\
The events were digitized and captured at the waveform level, resulting in 7 waveforms of 2.5 $\mu$s length for each event. A digital trapezoidal filter was applied offline to extract an estimate of the amplitude. The multiplicity information is extracted by looking at the number of channels for each event with an amplitude greater than a given threshold. Using only the events with multiplicity one, the energy spectrum can be constructed. \\
The energy calibration of the system was done using X-rays from a $^{55}$Fe and a $^{109}$Cd source.

\section{Measurements with a $^{109}$Cd commercial source}\label{sec3}
\begin{figure}
\centering
    \begin{subfigure}{}
        \includegraphics[width=0.4\textwidth]{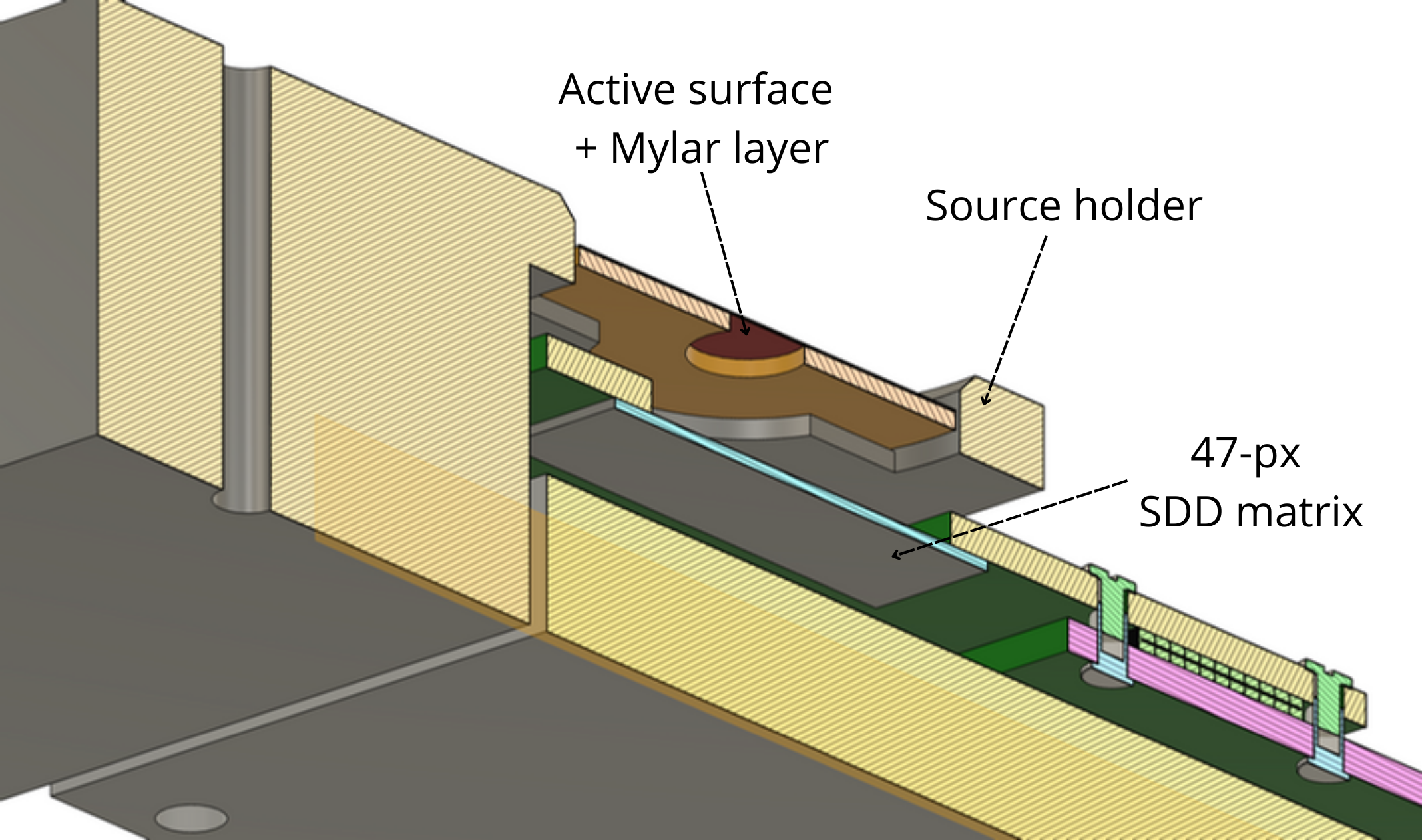}
    \end{subfigure}%
    \begin{subfigure}{}
        \includegraphics[width=0.4\textwidth]{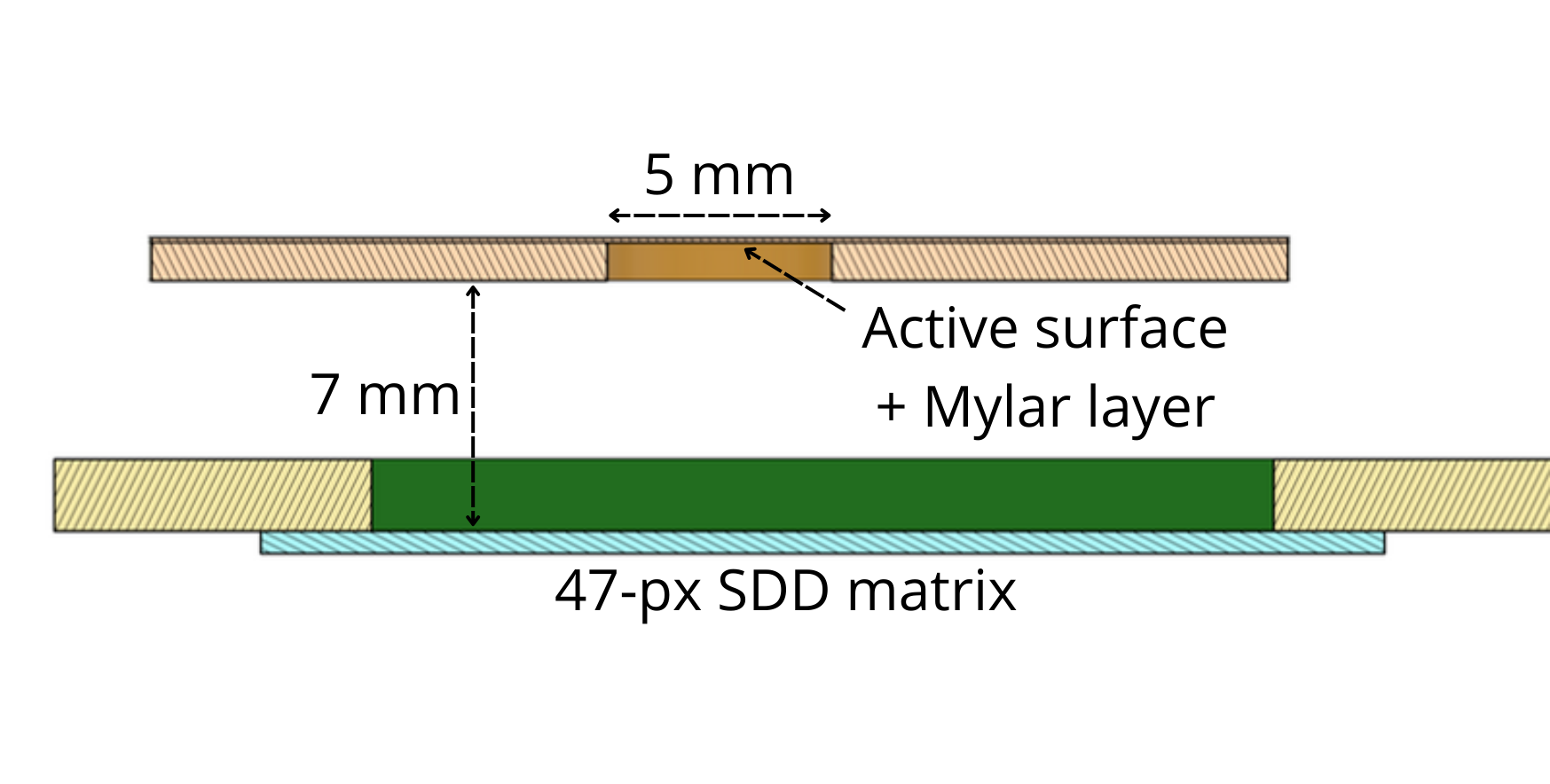}
    \end{subfigure}
\caption{CAD schematics of the setup used in the measurements. Section of source and detector (top), and side view (bottom).}\label{CAD}
\end{figure}

\begin{figure}[h]%
\centering
\includegraphics[width=0.5\textwidth]{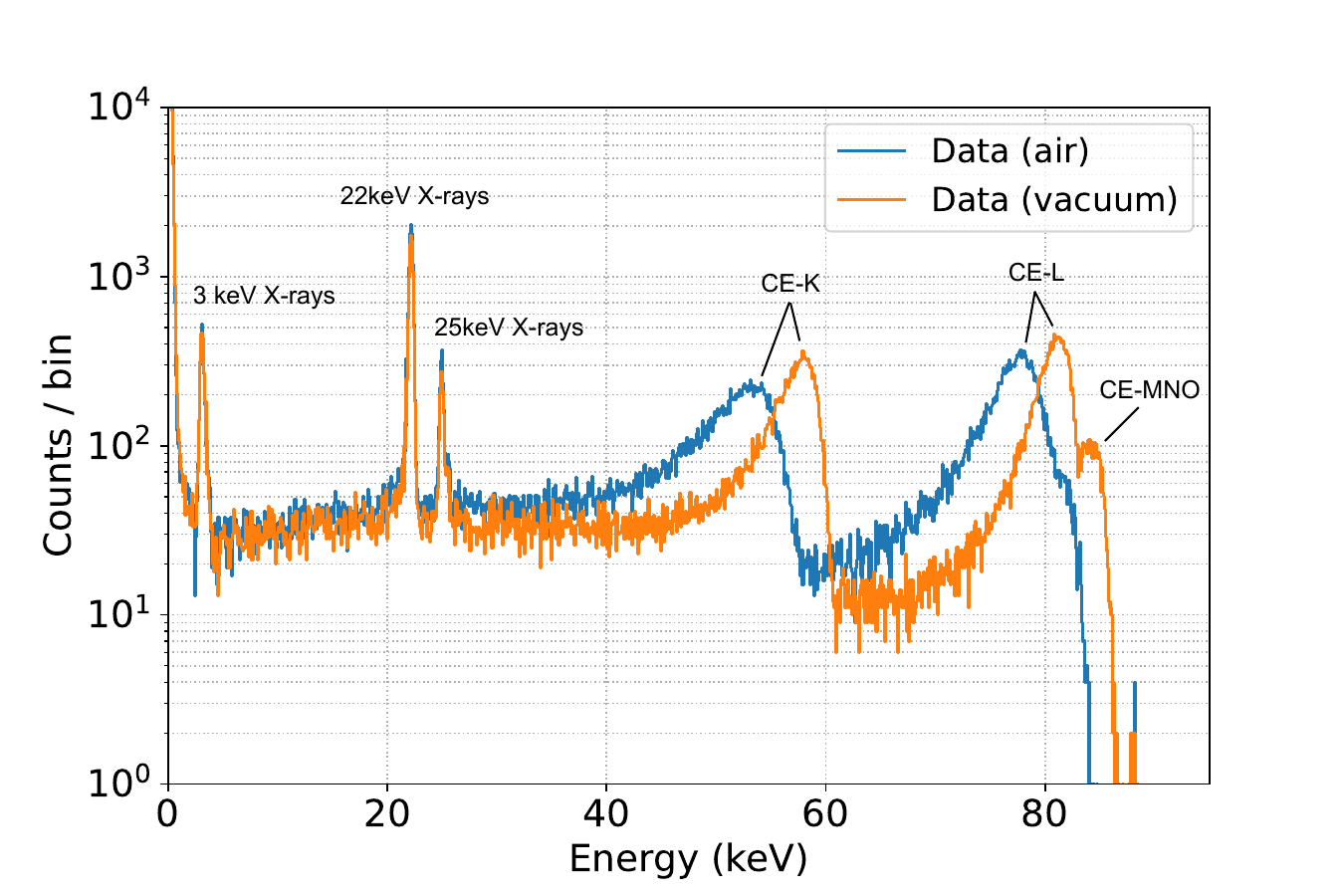}
\caption{Data acquired with SDD in a 3-hour measurement. The energy of X-ray peaks and the tag of the different IC electrons are shown.}\label{data_Cd}
\end{figure}

\begin{figure}[h]
    \begin{subfigure}{}
        \includegraphics[width=0.5\textwidth]{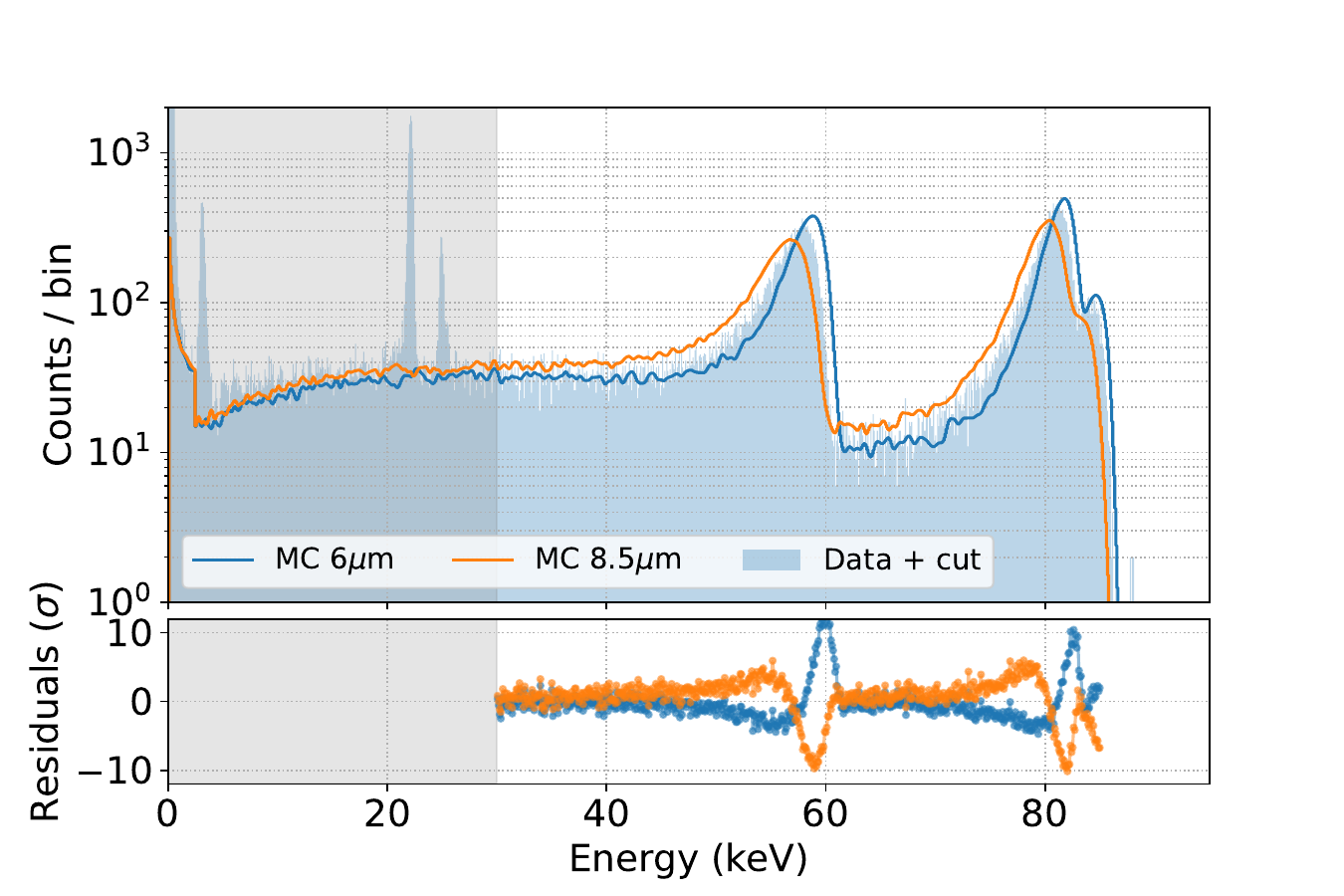}
    \end{subfigure}%
    \begin{subfigure}{}
        \includegraphics[width=0.35\textwidth]{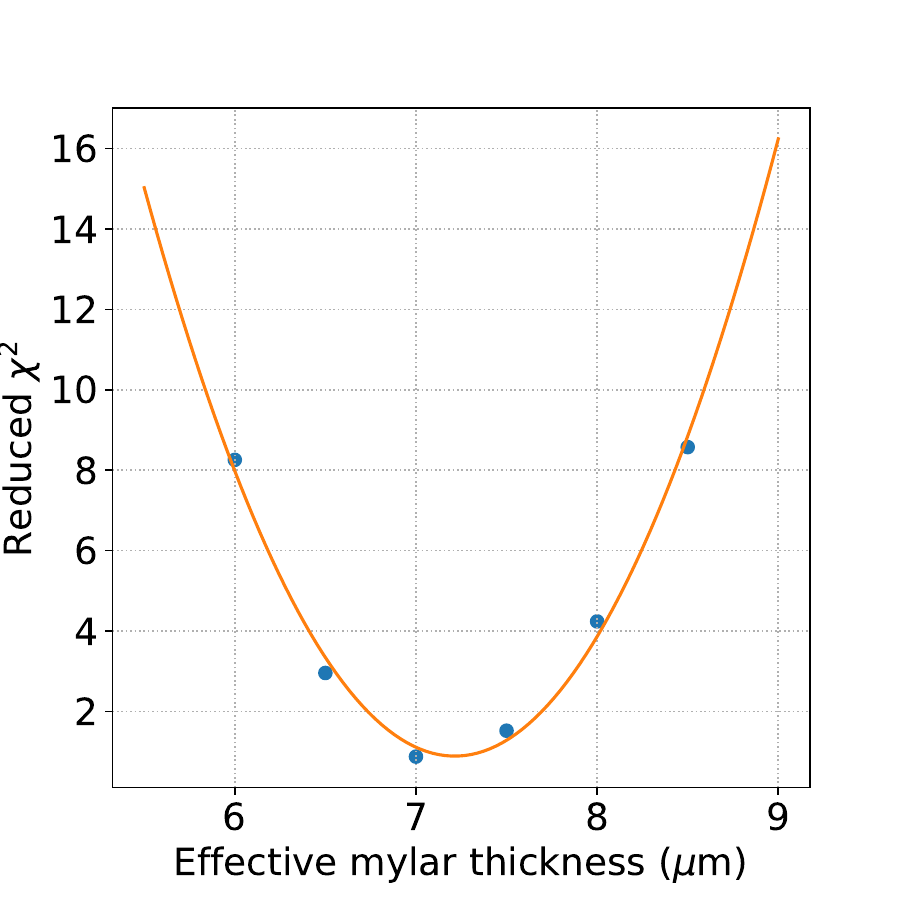}
    \end{subfigure}
\caption{Data-MC comparison for different values of the effective Mylar thickness (top). Reduced $\chi^{2}$ as a function of the effective Mylar thickness (bottom). The best-fit value is extracted through a parabolic fit.}\label{model_Cd}
\end{figure}

\begin{figure}[h]%
\centering
\includegraphics[width=0.5\textwidth]{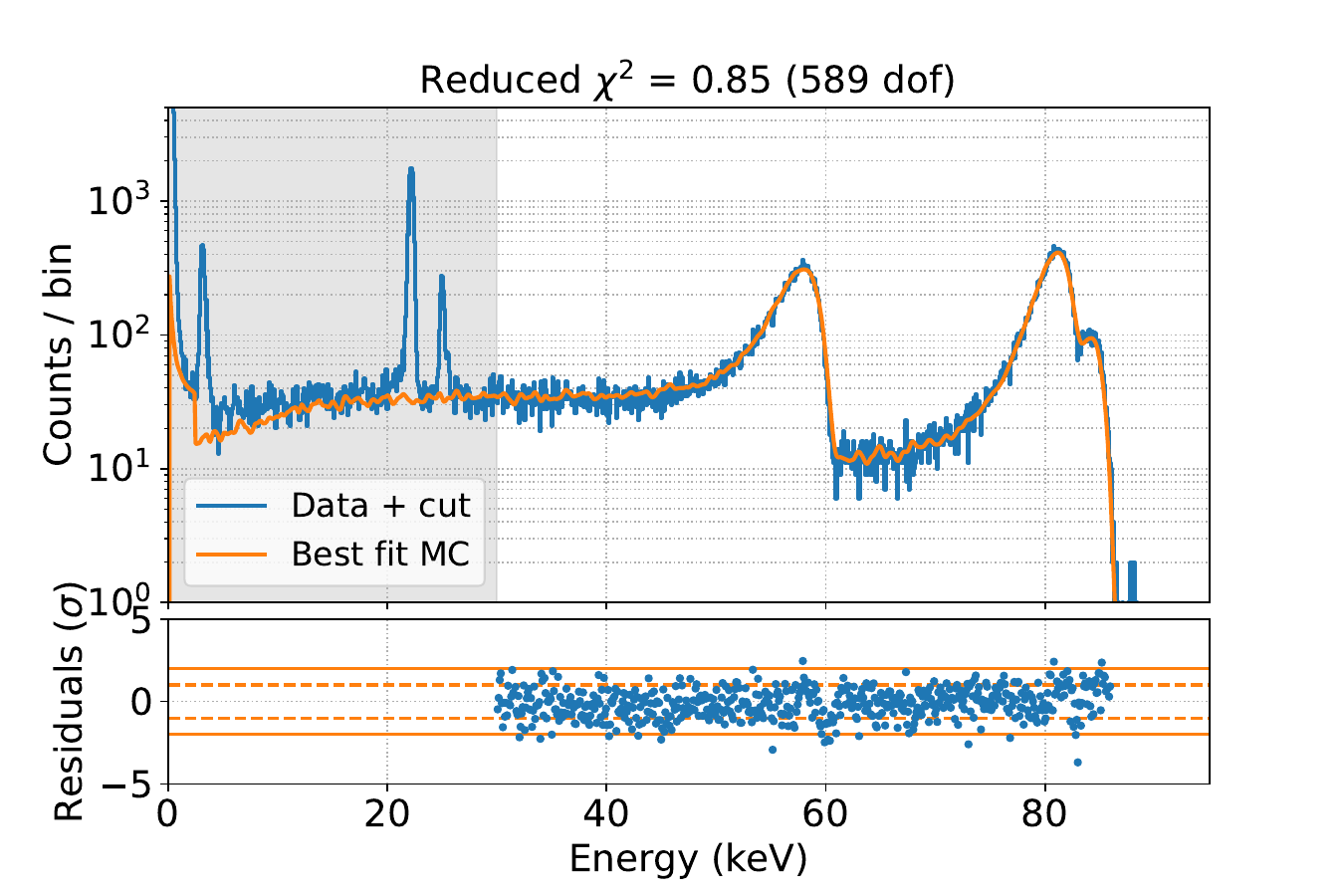}
\caption{Best fit of MC prediction to the data set acquired with the $^{109}$Cd source in vacuum. The fit is done only in the non-shaded area.}\label{fit_Cd_vacuum}
\end{figure}

\begin{figure}[h]%
\centering
\includegraphics[width=0.5\textwidth]{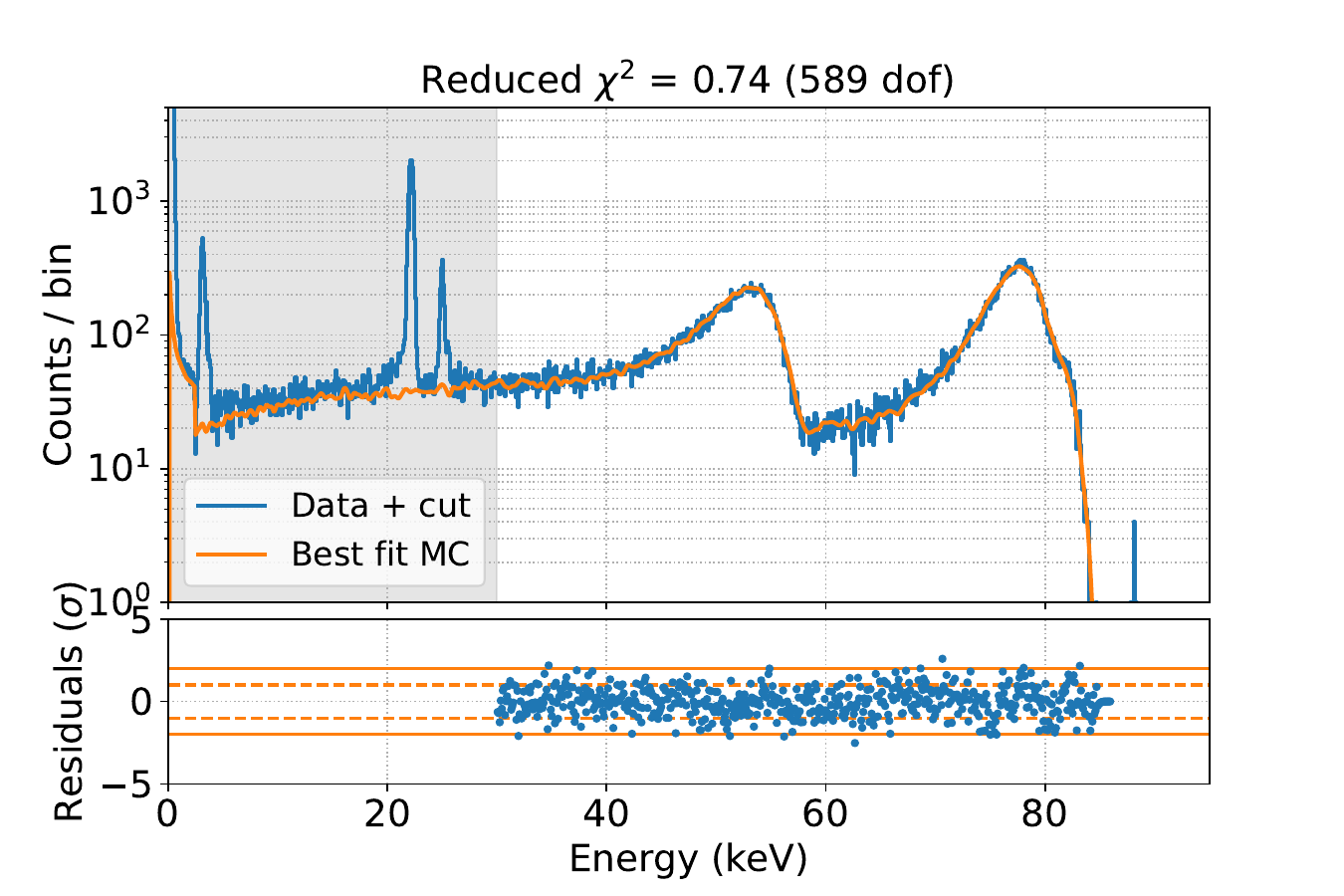}
\caption{Comparison of MC prediction to the dataset acquired with the $^{109}$Cd source in air. The comparison is done only in the non-shaded area.}\label{fit_Cd_air}
\end{figure}

For the first measurements, we used a commercial $^{109}$Cd source encapsulated in two thin aluminized mylar foils (each $\sim$6.5 $\mu$m thick). \\
This source decays by EC to the $^{109m}$Ag metastable state at 88 keV with a half-life of 462.1 d and emits mainly low energy X-rays at $\sim$3 keV, $\sim$22 keV and $\sim$25 keV. The $^{109m}$Ag decays to the ground state mainly by IC (B.R. $\sim$96.3$\%$) and has a half-life of 39.6 s. The IC electrons are emitted with different energies depending on the atomic shell they come from (see \ref{Tab.1}).
\begin{table}[h]
\begin{tabular}{c c c}
\toprule
Shell & Electron energy (keV) & Intensity (\%)\\
\midrule
K    & 62.5  & 41.8 \\ 
L    & $\sim$ 84.4  & 44.1  \\
M-N-O    & 87-88  & 10.4 \\
\botrule
\end{tabular}
\caption{Table of $^{109}$Cd electron lines.}\label{Tab.1}
\end{table}
Due to the presence of the Mylar layers, the electrons leaving the source are already less energetic and non-monochromatic. The position of the peaks and their width in the collected energy spectrum depend on the thickness of the Mylar, the presence of air between the source and the detector, and the position of the source with respect to the SDD. To remove one degree of freedom from the simulation, a source holder was 3D printed to center the source with the main SDD at a fixed distance of 7 mm (as shown in CAD scheme \ref{CAD}). \\
Two measurements of 3 h each were performed, one in vacuum at a pressure of $10^{-5}$ mbar and one at atmospheric pressure. The two spectra are shown in figure \ref{data_Cd}.
The position of the X-ray peaks is the same in both vacuum and air, as expected, because the photons interact directly in the detector, while the electron peaks are shifted to lower energies, depending on the minimum energy loss in the path between the source and the SDD, which is higher in the presence of air. Scattering in air also leads to a broader peak. \\
Our goal is to build a model that can reproduce the shape of the spectrum measured for these mono-energetic electrons. Based on previous studies \cite{Nava2023}, we wrote a GEANT4 simulation using the Penelope physics list \cite{Asai2021}, a package optimized for low-energy electromagnetic physics. The main SDD is simulated as a hexagonal prism surrounded by six other silicon detectors, and its entrance window is segmented into 30 layers, following the procedure described in \cite{Nava2022} to apply dead-layer effects. \\
Energy resolution is then applied by convolution of the spectrum with a Gaussian. \\
It's worth noting that even with the M2 cut, there remains a small effect of partial charge collection at the SDD boundary. This is due to those events that produce a signal below the threshold in surrounding SDDs, resulting in spurious M1 events. This effect is included in the model by assuming a Gaussian charge cloud in silicon, whose width is assumed to be energy independent and fixed at 15 $\mu$m from previous characterizations. \\
Electrons are then generated isotropically between the two Mylar sheets. To compare the simulation with the data, we expect the largest systematic effect to be the scattering in the Mylar, since the thickness of this part is $\sim500$ times larger than the detector entrance window. In addition, the peak broadening caused by multi-scattering in the source is larger than the energy resolution of the SDD. Therefore, we decided to fix the entrance window parameters and the energy resolution to values obtained from previous characterization measurements (see \cite{Biassoni2020}). \\
We ran simulations for different thicknesses of these layers to give the simulation enough freedom to compensate for other unsimulated materials, such as the small amount of aluminum or very thin adhesive layers used in the source fabrication, but whose thickness is not specified. Thus, the main free parameter we use is an effective thickness of the Mylar sheets. The best-fit estimate for the effective Mylar thickness is the one that minimizes the $\chi^2$ between the MC prediction and the data. Since the measured electrons are at higher energies than the X-rays used to calibrate, two additional nuisance parameters (a horizontal gain and shift) are left free in the $\chi^2$ minimization. For all the fits the best estimation for these parameters is compatible with the calibration made using X-rays. \\
Figure \ref{model_Cd} shows that different Mylar thicknesses cause a shift and a different broadening of the main $^{109}$Cd IC electron peaks.
The data-MC comparison is performed in the energy region $>$30 keV to avoid contributions from unsimulated Ag X-rays, for five different values of the effective Mylar thickness. A parabolic fit of the $\chi^2$ as a function of the effective Mylar thickness yields a best-fit estimate of 7.2 $\mu$m for this parameter, as shown in Figure \ref{model_Cd}. The corresponding spectrum is shown in Figure \ref{fit_Cd_vacuum}. \\
The fit result is in excellent agreement with the data, and in particular, all major structures in the spectrum are well reproduced, from the positions of the peaks to their widths and the shape of their tails. \\
After this step, we decided to fix the best-fit result for the Mylar thickness and try to reproduce the air data set without adding any new free parameters, as a cross-check of GEANT4's ability to predict the effect of additional material interposed between the source and the detector. The comparison is shown in figure \ref{fit_Cd_air}.
The still excellent reproduction of the data proves the reliability of the model and its robustness to changes in the experimental conditions.

\section{Measurements with $^{14}$C}\label{sec4}
\begin{figure}[h]
\centering
    \begin{subfigure}{}
        \includegraphics[width=0.5\textwidth]{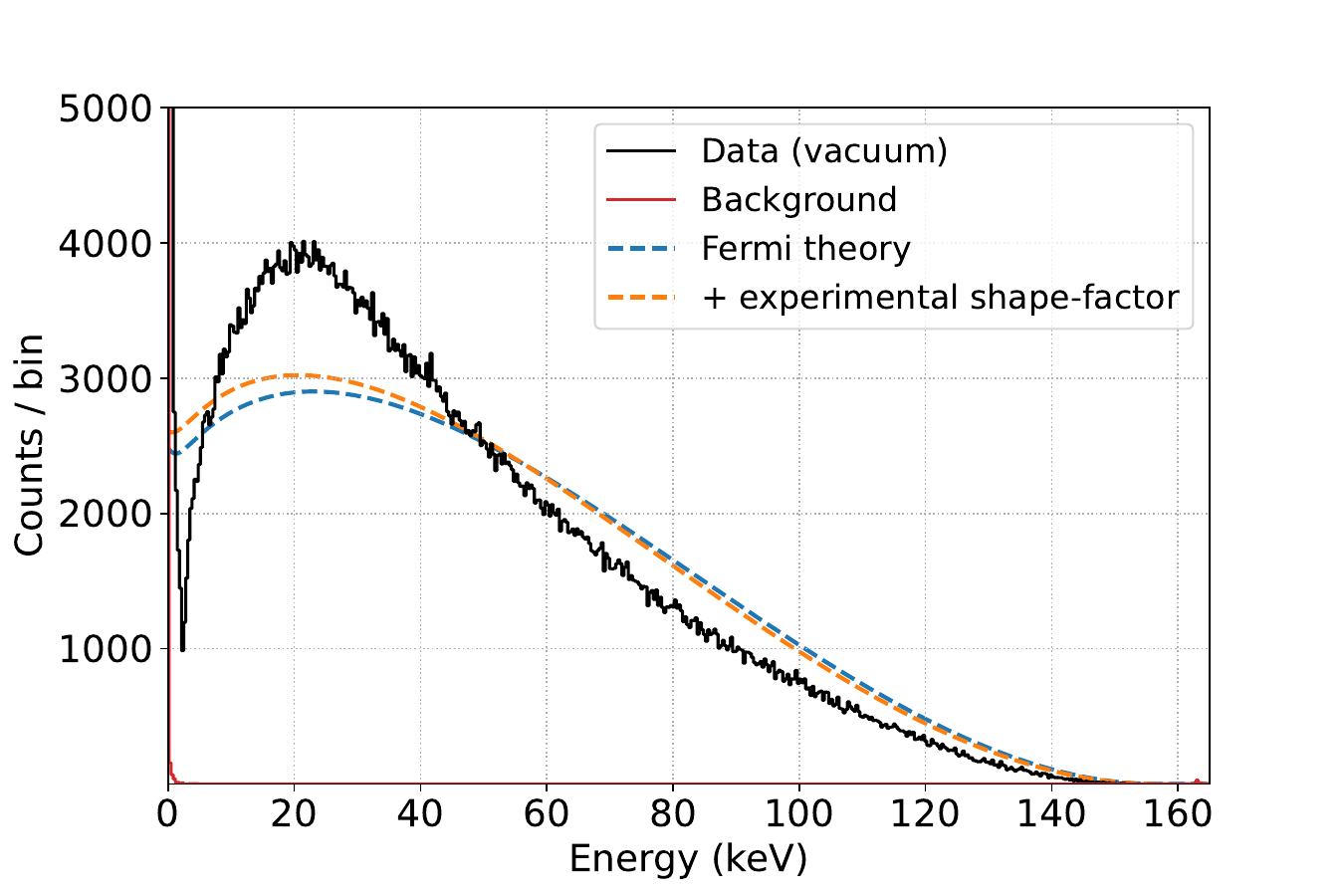}
    \end{subfigure}%
    \begin{subfigure}{}
        \includegraphics[width=0.35\textwidth]{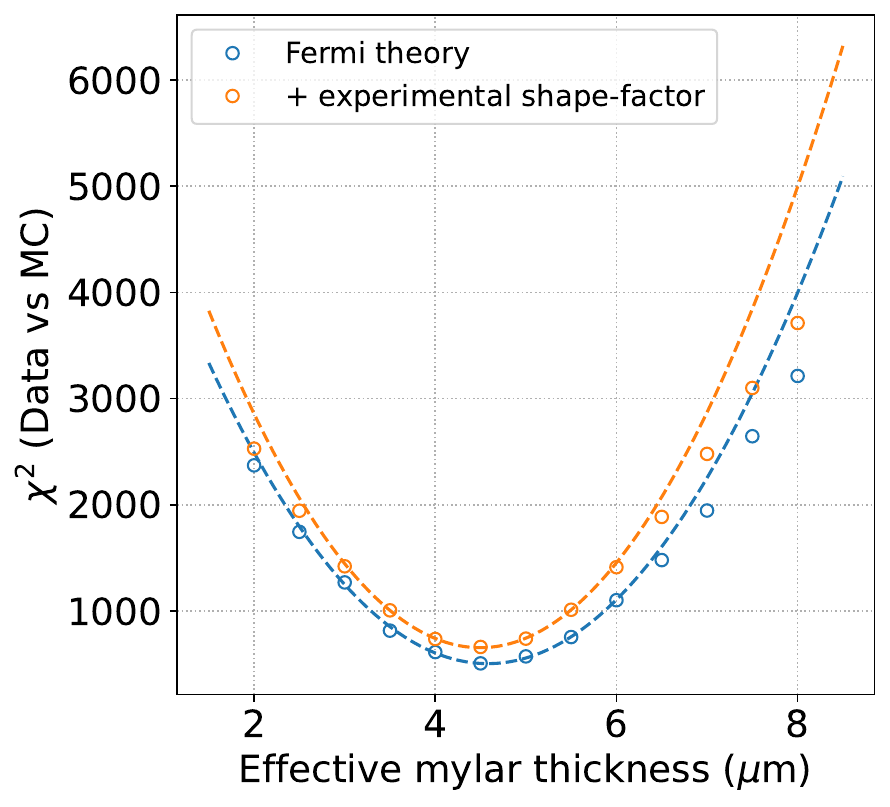}
    \end{subfigure}
    \caption{Data acquired with an SDD in a 6 h measurement using a $^{14}$C source (top). $\chi^2$ as a function of the effective Mylar thickness for the two models: the Fermi theory prediction and the one including the experimental shape factor (bottom).}\label{data_C}
\end{figure}

\begin{figure}[h]%
\centering
\includegraphics[width=0.5\textwidth]{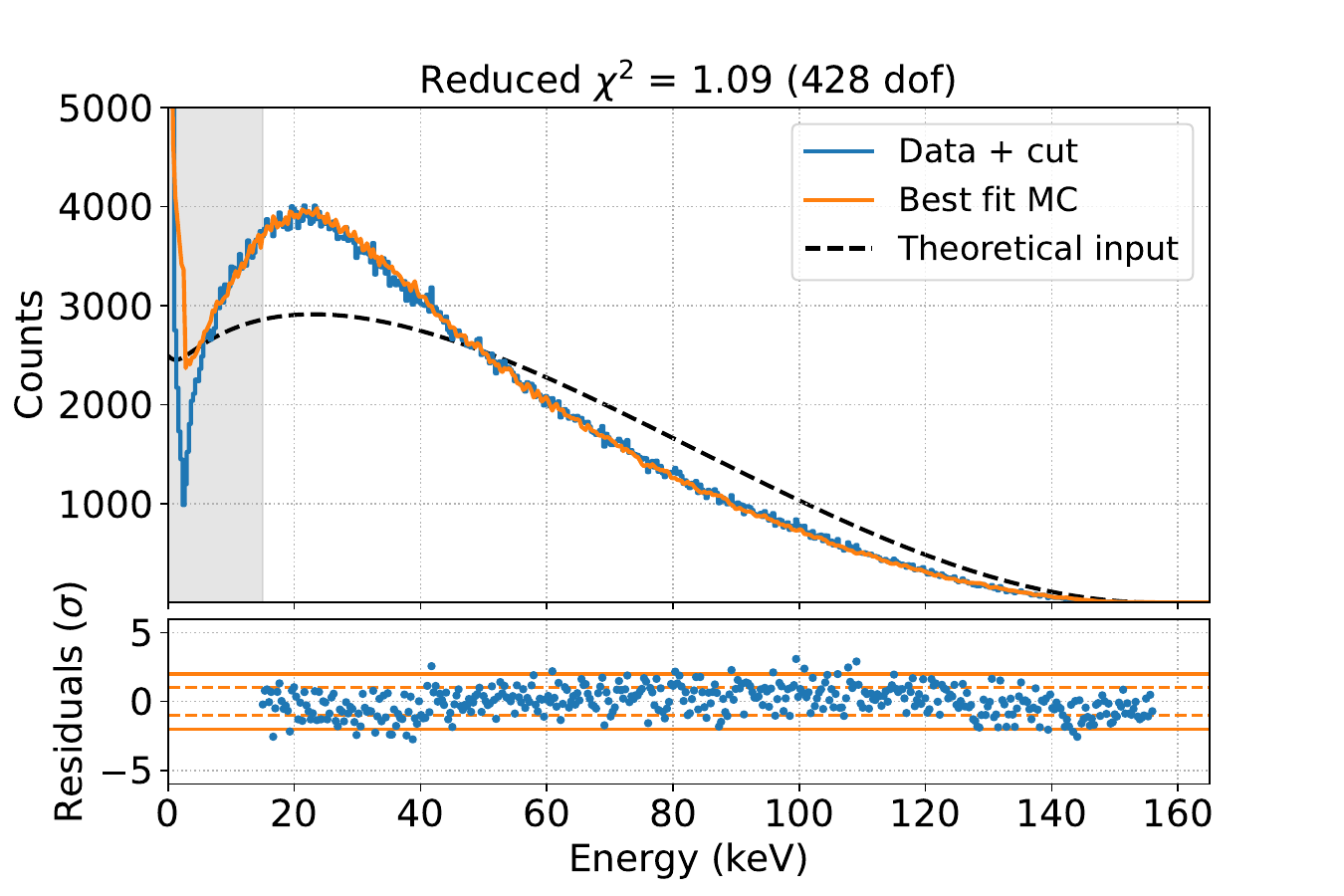}
\caption{Fit to the data set acquired with the $^{14}$C source in vacuum using the MC prediction. The fit is performed only in the non-shaded area. The theoretical input is also shown.}\label{fit_C}
\end{figure}

\begin{figure}[h!]
\centering
\begin{subfigure}{}
\includegraphics[width=0.5\textwidth]{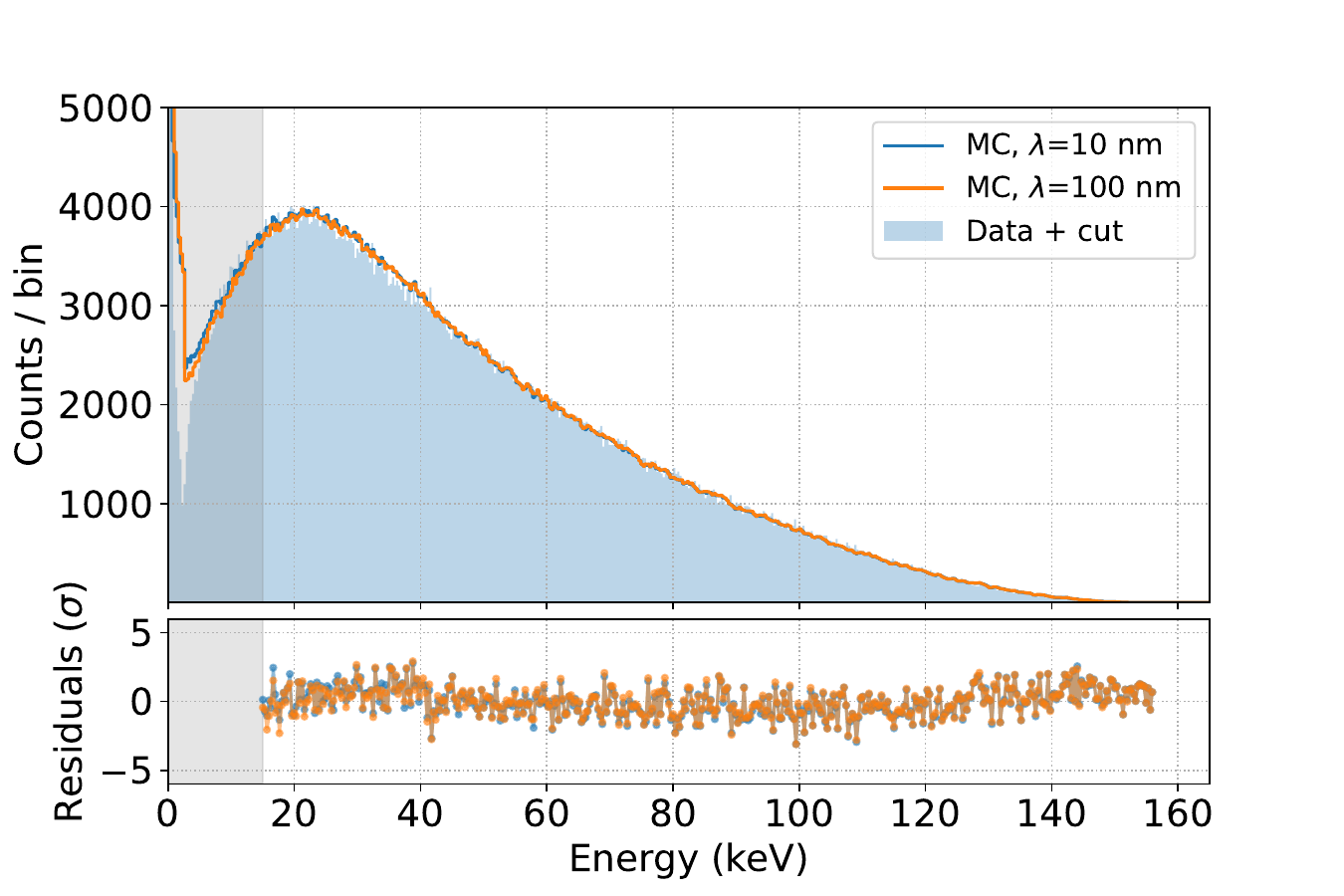}
\end{subfigure}
\begin{subfigure}{}
\includegraphics[width=0.5\textwidth]{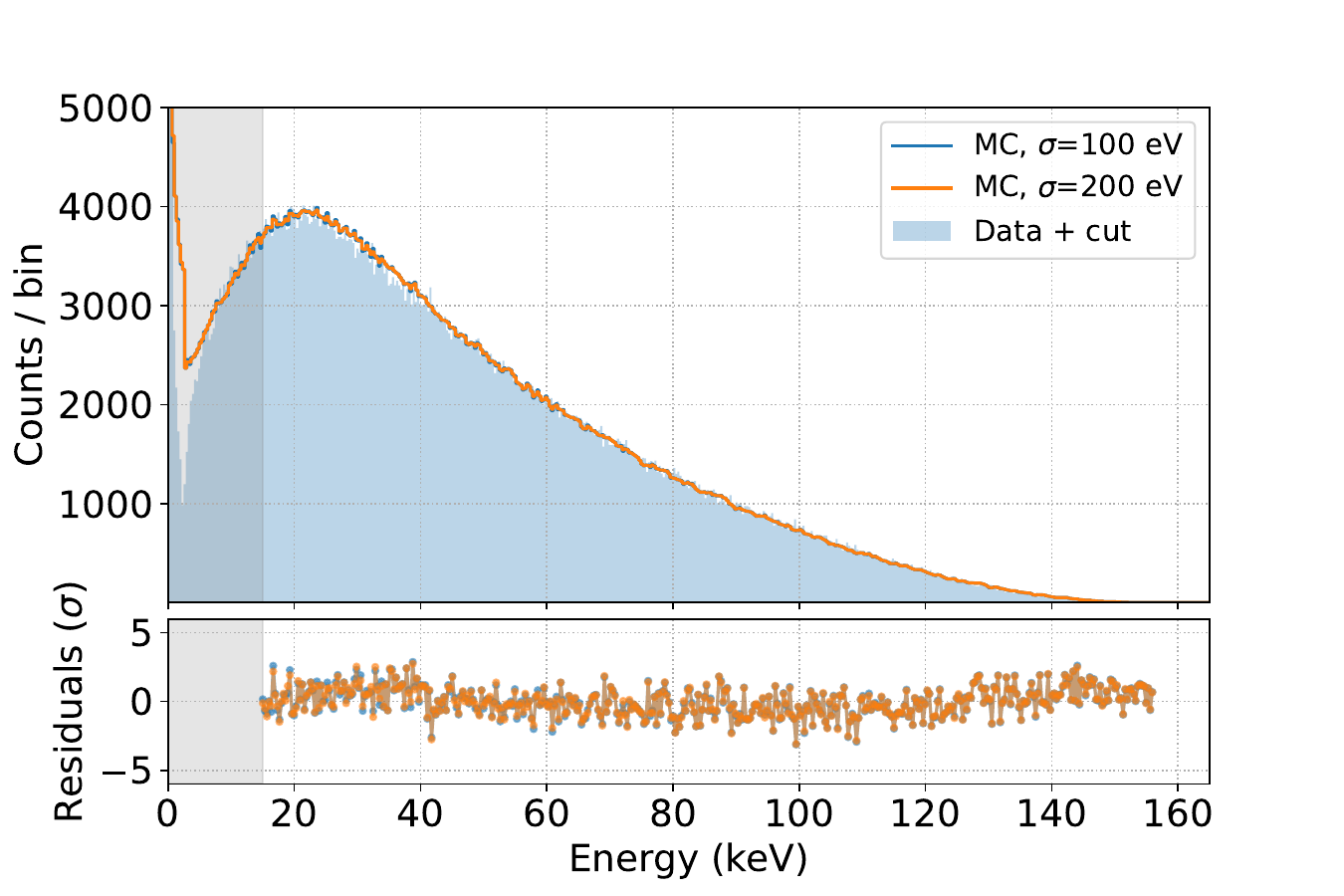}
\end{subfigure}
\begin{subfigure}{}
\includegraphics[width=0.5\textwidth]{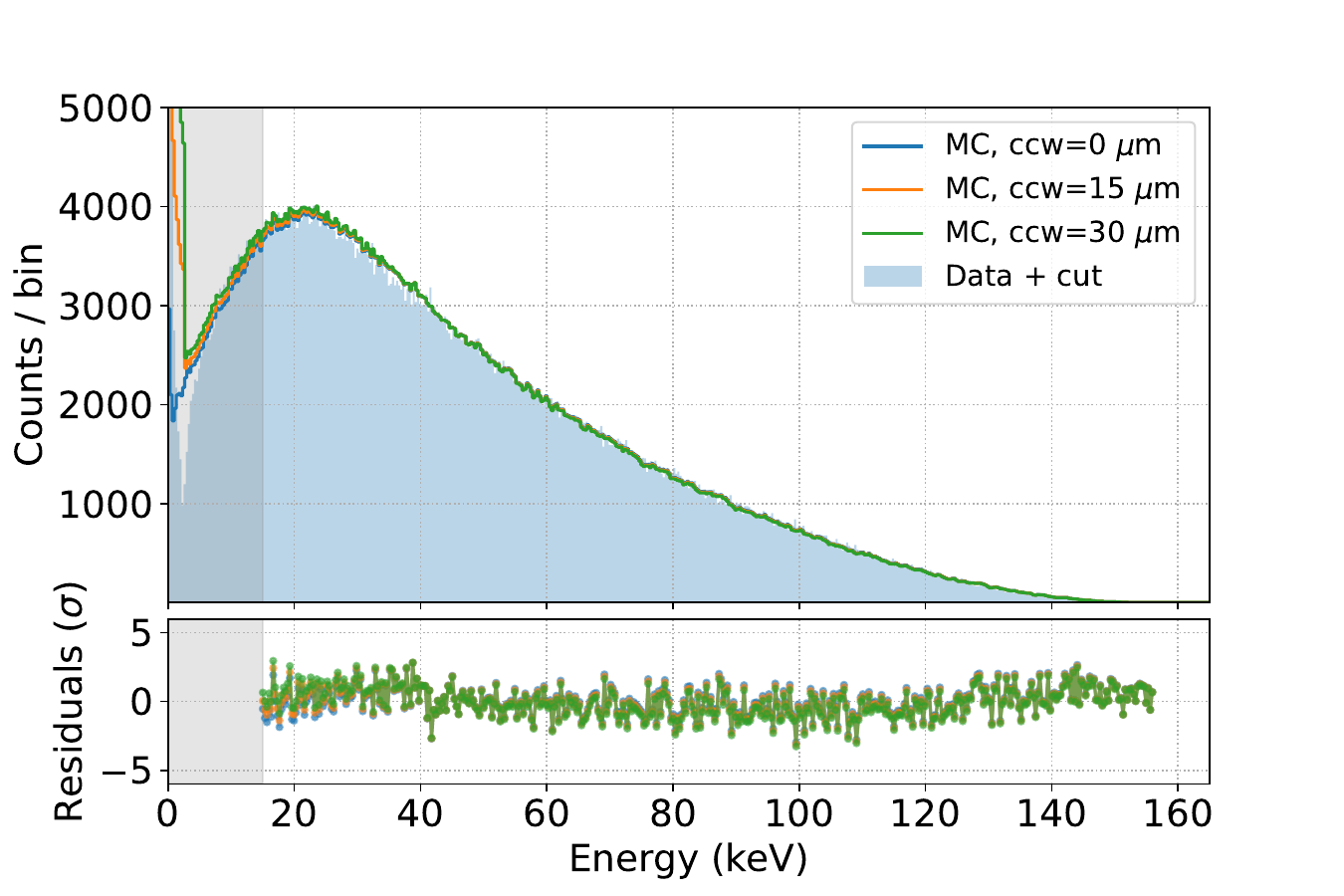}
\end{subfigure}
\caption{Fits obtained by varying $\lambda$ (top), the baseline resolution $\sigma$ (center) and the charge cloud width in Si (bottom).}\label{test_det_eff}
\end{figure}

\begin{figure}[h!]
\centering
\begin{subfigure}{}
\includegraphics[width=0.5\textwidth]{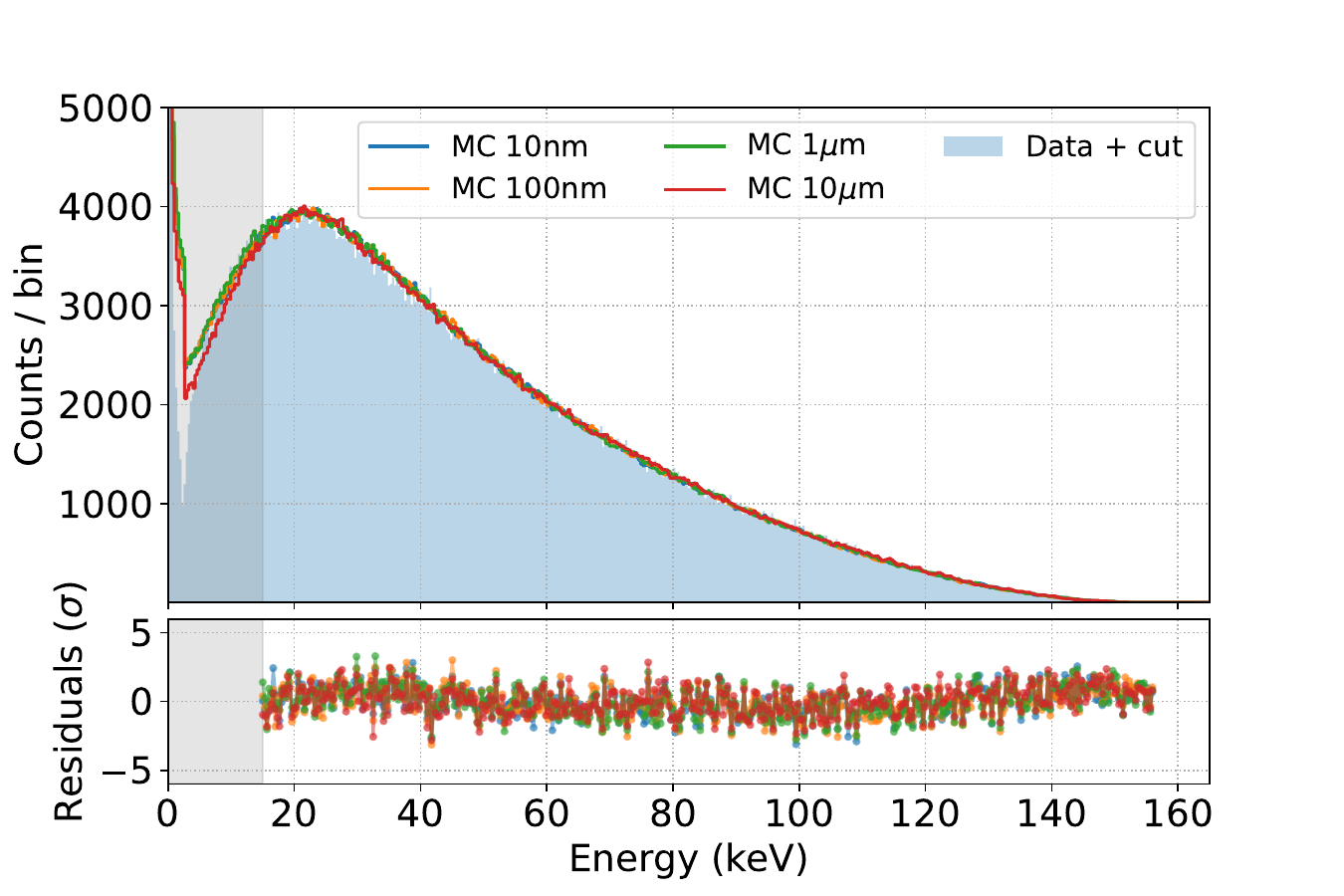}
\end{subfigure}
\begin{subfigure}{}
\includegraphics[width=0.5\textwidth]{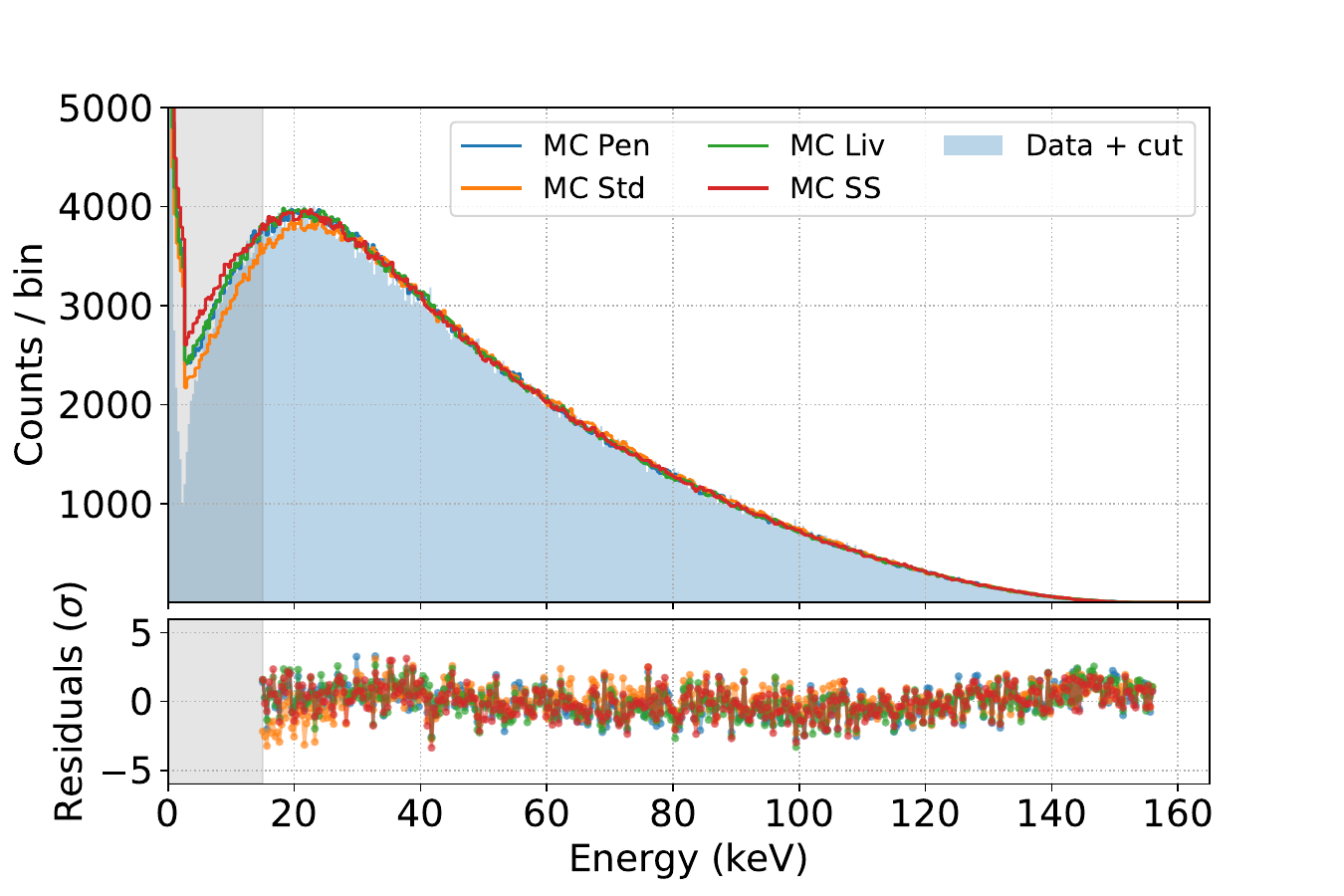}
\end{subfigure}
\caption{Fits obtained by varying the production cut for secondaries in GEANT4 (top) or the physics list used  (bottom) in the simulations.}\label{test_G4_eff}
\end{figure}

We then switched to a commercial $^{14}$C source encapsulated between a 100 $\mu$m thick paper foil on the back and a thin aluminized Mylar layer on the front. \\
The $^{14}$C allowed $\beta$ decay has a Q-value of $\sim$ 156 keV and an average electron energy of $\sim$ 50 keV. So we pushed our measurements to slightly higher electron energies. \\
We performed a 6 hour measurement using the same setup described in the previous section. The resulting spectrum, calibrated against the $^{109}$Cd peaks, is shown in Figure \ref{data_C}.
The theoretical spectral shape of $^{14}C$ can be obtained using software such as Betashape \cite{Mougeot2015}. In particular, two spectra are available: a purely theoretical prediction based on $\beta$ kinematics and the Fermi factor, and a prediction including an energy-dependent experimental shape factor measured in \cite{Kuzminov2000}. The shapes of the two predictions, normalized to the integral above 15 keV, are shown in Figure \ref{data_C}. These spectra were used as input for the GEANT4 simulations. \\
It is clear how the response of the system changes the shape of the spectrum, again highlighting the need for an accurate and reliable simulation to interpret the experimental measurement. \\
We performed the same analysis as for the $^{109}$Cd case for these two models, varying the effective Mylar thickness in the simulation. The result of the fit, performed starting at 15 keV, is shown in Figure \ref{data_C}. \\
In both cases, the best fit value for the effective Mylar thickness is 4.5 $\mu$m, indicating that the effect of this thickness is not degenerate with the shape factor. The model without the shape factor better describes the data for all values of the only free parameter (the Mylar thickness), with a $\Delta \chi^2 \sim 150$ in the minimum, indicating that Fermi theory is preferred and sufficient to explain the $^{14}$C spectrum. The best fit, without any experimental shape factor, is shown in Figure \ref{fit_C}.\\
We then tested the robustness of the model's prediction by comparing the results obtained by varying the parameters of the detector response, namely the $\lambda$ parameter related to the depth-dependent charge collection efficiency (default: $\lambda$ = 55 nm \cite{Biassoni2020}), the baseline energy resolution (default: $\sigma$ = 150 eV), and the charge cloud width in Silicon, which mimics the partial charge collection at the SSD border (default: ccw = 15 $\mu$m). The results are shown in figure \ref{test_det_eff}.\\
The effect on the fit quality of using a ten times larger $\lambda$ parameter is negligible, as is the effect of using a resolution 50 eV higher or lower than the reference. This is again due to the fact that the scattering within the source dominates the broadening of the measured spectra. The effect of a zero or doubled charge cloud with respect to the standard results in only a small change in the low energy part of the spectrum, where the multiplicity cut is less efficient. We can therefore conclude that the spectrum above 15 keV is almost independent of the parameters of the SDD response. \\
To assess the robustness of our prediction, we also repeated the comparison by varying some settings of the GEANT4 simulation, namely the secondary production cut (the distance that secondary particles have to travel in a given material to be produced in GEANT4) and the physics list (the set of physical models used from the simulation)\cite{Ivanchenko2011}. The results are shown in the figure \ref{test_G4_eff}.\
We see an effect in the low energy region only for production cuts as high as 10 $\mu$m, while the default for our simulations is 10 nm. \\
Regarding the available physics lists, The Standard Electromagnetic \cite{Ivanchenko2011} is the only one that produces worse results in the low energy region of the spectrum, while the physics lists specifically designed for low energy electromagnetic interactions, such as Penelope, Livermore, and Single Scattering \cite{Ivanchenko2011}, produce the same result in the region of interest. As mentioned above, we used Penelope as the reference for all simulations. \\

\section{Conclusions and outlook}\label{sec5}
In this work we have demonstrated the capability of SDDs to measure electrons in the 15 - 150 keV range using $^{109}$Cd and $^{14}$C sources. The need for an accurate model of the system response has been emphasized, and in this context the performance of GEANT4 low-energy simulations in reconstructing the measured spectra has also been shown. The settings of the simulation and the knowledge of the detector response are of minor importance as long as we use commercial sources encapsulated in thin passive layers. Instead, the most important systematic effect is the exact knowledge of the thickness of this layer. By varying the effective thickness, we were able to obtain an excellent reconstruction of the shape of the $^{109}$Cd monochromatic lines, both in vacuum and in air, and of the $^{14}$C $\beta$ spectrum. Other measurements found in the literature (\cite{Kuzminov2000}) required the introduction of an experimental shape factor to explain the shape of the measured spectrum, while in our case this requirement is rejected with high significance. \\
With these measurements, we have therefore validated a new SDD-based technique for measuring $\beta$ spectra, which is important for its complementarity with respect to cryogenic calorimetric measurements and for its versatility in source selection. \\ 
In the near future we will switch to a new detector technology using larger and thicker SDDs. The side-length will be $\sim$1 cm, leading to a smaller fraction of events near the borders, and therefore to an even smaller systematics related to partial charge collection. The thickness will be 1 mm, allowing us to measure electrons with higher energies that would otherwise not be fully contained. The first decay we will study is the non-unique second forbidden decay from a commercial $^{99}$Tc source. \\
We are also planning to improve the source-related systematics by actually depositing the radioactive material on an auxiliary detector, to avoid a passive layer between the source and the main SDD, and to allow anti-coincidence measurements, by which those events with only partial energy deposition in the main SDD can be vetoed. \\
A final consideration concerns the possibility of studying the half-lives of isotopes: in this work we have carried out a shape-only analysis of the spectra, but when switching to custom deposited sources, through a precise knowledge of the amount of the isotope of interest, half-life measurements will become possible, although still challenging. \\

\section*{Acknowledgments}
We acknowledge financial support under the National Recovery and Resilience Plan (NRRP), Mission 4, Component 2, Investment 1.1, Call for tender No. 104 published on 2.2.2022 by the Italian Ministry of University and Research (MUR), funded by the European Union – NextGenerationEU– Project Title ASPECT-BET: An sdd-SPECTrometer for BETa decay studies – CUP H53D23001020006 - Grant Assignment Decree No. 974 adopted on June 30, 2023 by the Italian Ministry of Ministry of University and Research (MUR). \\
We thank the KATRIN collaboration and especially the TRISTAN working group for providing the TRISTAN detector prototype used in these measurements.


\bibliography{main}

\end{document}